\documentclass[fleqn,10pt]{SelfArx} % Document font size and equations flushed left

\usepackage[english]{babel} % Specify a different language here - english by default

\usepackage{lipsum} % Required to insert dummy text. To be removed otherwise

\usepackage{tabularx}
\usepackage{amsmath}
\usepackage{natbib} 
\usepackage{siunitx}
\usepackage{fixmath}
\usepackage{booktabs}
\usepackage{makecell}
\usepackage[hyphens]{url}
\usepackage{flushend}

\definecolor{color1}{RGB}{0,0,90} % Color of the article title and sections
\definecolor{color2}{RGB}{0,20,20} % Color of the boxes behind the abstract and headings

\usepackage{hyperref} % Required for hyperlinks

\hypersetup{
	hidelinks,
	colorlinks,
	breaklinks=true,
	urlcolor=color2,
	citecolor=color1,
	linkcolor=color1,
	bookmarksopen=false,
	pdftitle={Title},
	pdfauthor={Author},
}

\JournalInfo{Preprint} % Journal information
\Archive{} % Additional notes (e.g. copyright, DOI, review/research article)

\PaperTitle{College Closures from 2020 to 2025:\\ An Exploratory Analysis and Its Implications for the Enrollment Cliff} % Article title

\PageTitle{College Closures from 2020 to 2025}
\JournalName{Preprint}

\Authors{Sepehr Akbari\textsuperscript{$\dagger$}, Shahrzad Jamshidi\textsuperscript{$\dagger$}\textsuperscript{$\ddagger$}} % Authors
\affiliation{\textsuperscript{$\dagger$}\textit{Department of Mathematics and Computer Science, Lake Forest College, USA}} % Author affiliation
\affiliation{\textsuperscript{$\ddagger$}\textbf{Corresponding author}: sjamshidi@lakeforest.edu} % Corresponding author
 
\Keywords{College Closures~--~Higher Education~--~Enrollment Cliff~--~Institutional Vulnerability}
\newcommand{\keywordname}{Keywords} % Defines the keywords heading name

\Abstract{The COVID-19 pandemic produced a modest wave of college closures and mergers that may offer an early, if imperfect, preview of the demographic ``enrollment cliff'' anticipated in the coming decade. This paper examines the institutions that closed or merged between 2020 and the end of 2025. We assemble a dataset of 65 such institutions, pairing institutional characteristics with state- and regional-level demographic, economic, and financial indicators, and supplement it with a corpus of news coverage of the closures. Using a combination of Bayesian models, dimensionality reduction, clustering, and topic modeling, we describe where these closures occurred, what the closed institutions had in common, and how they were discussed publicly. Consistent with prior demographic projections, closures were more frequent in the Northeast and Midwest, though the absolute numbers remain small. The closed institutions were heterogeneous rather than uniform: financial structure, regional demographics, and institutional mission each contributed to distinguishing them, and religious affiliation recurred prominently in media coverage. We frame these results as exploratory and descriptive given the small sample, and we discuss what they may, and may not, imply for institutions navigating the enrollment cliff.}
 
\begin{document}

\maketitle

\thispagestyle{empty}

%----------------------------------------------------------------------------------------
%	ARTICLE CONTENTS
%----------------------------------------------------------------------------------------

\section{Introduction}

\subsection{Background}

The {\it enrollment cliff} refers to the projected decline in U.S. college and university enrollments, driven by demographic changes---specifically, declining birth rates during and after the Great Recession (2007--2009). Historically, the annual share of women aged 15--44 who gave birth remained stable between 6.5\% and 7\% from 1980 to 2007. Since the Great Recession, however, this rate has steadily declined to 5.6\%, reflecting a 14\% to 20\% reduction in births between 2007 and 2020. \citet{Kearney2022} document that this trend spans diverse demographic groups, with women having fewer children and at older ages. Although they studied several potential drivers, no single factor emerges as the primary driver of state-level differences in birth rates. The persistence of similarly low birth rates in other high-income countries further suggests that systemic shifts, such as changing social norms or economic uncertainty, may be reshaping fertility patterns, rather than localized conditions.  

Around 2025, the cohort born during this period of declining fertility began to reach traditional college age, precipitating a drop in the domestic college-aged population. This population shift poses challenges for higher education institutions, which must adapt to a landscape marked by fewer traditional domestic students. 

\begin{figure}[htbp]
    \centering
    \includegraphics[width=\linewidth]{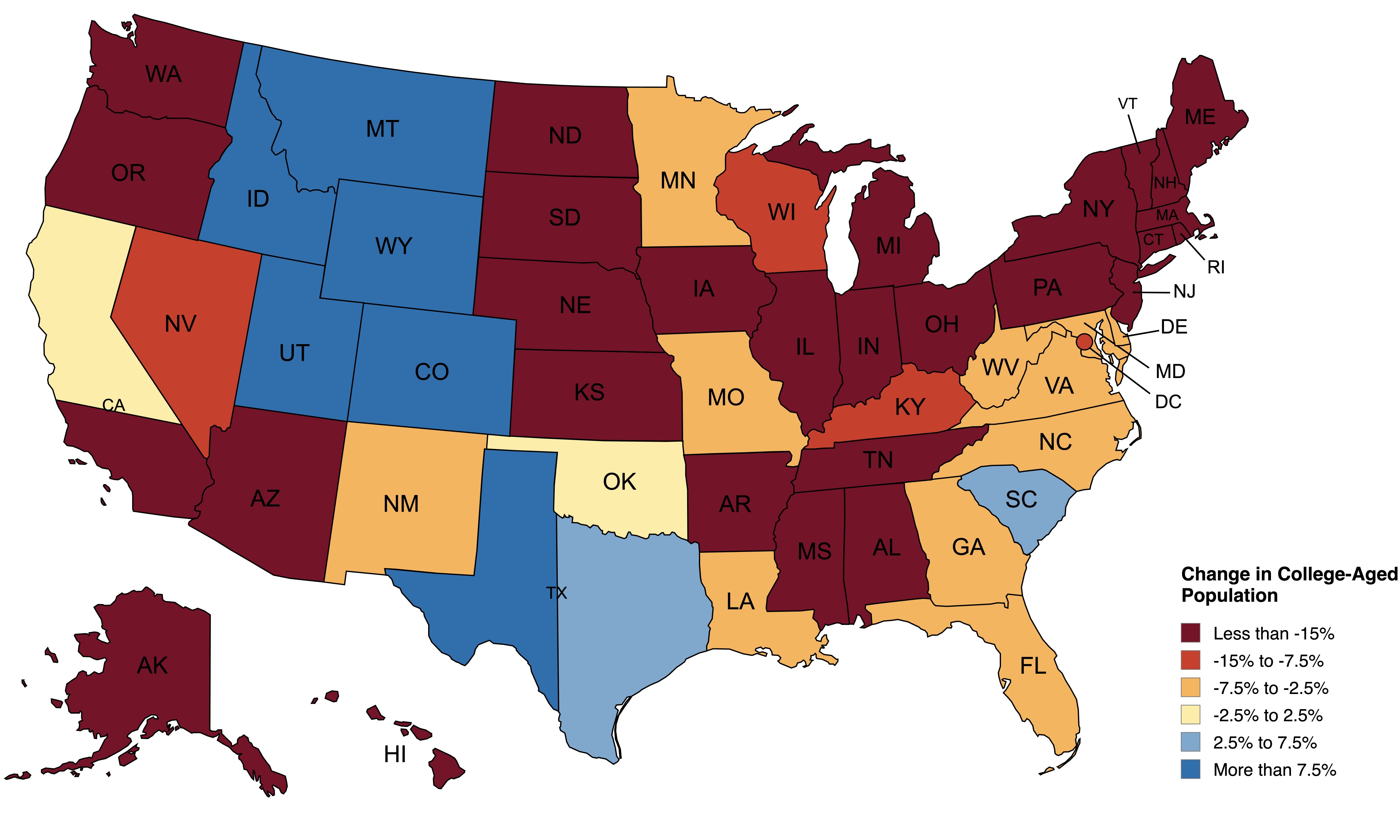}
    \caption{\footnotesize Predicted state-level changes in the college-age population across the United States from 2012 to 2029, with colors indicating projected losses or gains, based on demographic projections presented in \cite{Grawe2018-Book}. California and Texas were split into North/South and East/West, respectively, due to their size.} 
    \label{fig:popmap}
\end{figure}

In assessing the impact of the enrollment cliff, \citet{Grawe2018-Book} introduces the Higher Education Demand Index (HEDI) to estimate how demographic trends might influence future enrollments. Grawe's predictions are built from regression models that use the projected populations of 18-year-olds and the proportions of children expected to attend college, categorized by year and spanning 63 locations across the United States. His assessment anticipates a disproportionate impact on institutions in the Northeast and Midwest (traditional hubs of higher education), while regions such as the Southwest may experience relative stability due to migration and population growth. Figure \ref{fig:popmap} illustrates these projected regional disparities in college-aged populations, based on data in \citet{Grawe2018-Book}. 

The COVID-19 pandemic offers a partial preview of institutional vulnerability to enrollment changes and financial shocks. \cite{NSC2022} estimates that total postsecondary enrollment fell by almost 938,000 students nationally, a 5.1\% decline, from Fall 2019 to Fall 2021. \cite{FRBP2021} estimates that institutions of higher education faced \$70--\$115 billion in revenue losses. The pandemic's abrupt enrollment and financial disruptions are not equivalent to the slower demographic decline of the enrollment cliff, but the closures they produced offer one of the few recent opportunities to examine which institutions prove most vulnerable when enrollment and revenue contract.

It is worth being explicit about how these two phenomena differ, because that difference governs what the present study can and cannot show. The pandemic was an acute, broad shock: enrollment and revenue fell sharply across many institutions at roughly the same time, and the institutions that closed were often those least able to absorb a sudden loss of liquidity. The enrollment cliff, by contrast, is a gradual and geographically concentrated decline in the traditional college-aged population that erodes tuition demand over years rather than months. The two need not threaten the same institutions: a college with thin cash reserves might fail under the pandemic shock regardless of its demographic exposure, while a college in a region of declining births might weather the pandemic yet face sustained pressure once the cliff arrives. Where the two overlap---most plausibly among small, tuition-dependent institutions in regions of demographic decline---the pandemic-era closures offer a useful preview. We therefore treat the 2020--2025 closures as an imperfect analogue to, rather than a realized instance of, the enrollment cliff, and we interpret any resemblance between the two in that light.

\subsection{College Closure and Institutional Vulnerability}

Separate from demographic projections, a longstanding literature examines college closure, merger, and financial distress as related but not identical forms of institutional vulnerability. Early work developed accounting-based models of distress for private colleges \citep{Schipper1977-FinancialDistress} and linked aggregate changes in private-college closures and mergers to the balance between tuition revenue and operating costs \citep{BatesSanterre2000-ClosuresMergers}. Using a survival model for private institutions observed from 1975 to 2005, \citet{PorterRamirez2009-WhyCollegesFail} found that institutional scale, selectivity, endowment per student, and religious affiliation were associated with survival. Their results did not support baccalaureate emphasis or single-sex status as independent explanations once other characteristics were considered. These studies also make clear that merger and closure should not be treated as mechanically interchangeable: a merger may preserve instruction or institutional functions that an outright closure ends.

Later studies broaden the account beyond a generic image of the small private liberal-arts college. Most of the resource-constrained private colleges revisited by \citet{Tarrant2018-InvisibleColleges} continued to operate in some form decades after first being identified as vulnerable, an important counterweight to claims of inevitable failure. Sector- and mission-specific analyses also produce different correlates of persistence. Among HBCUs, urban location, remedial programs, and greater student-services spending were associated with continued operation in \citet{Britton2023-Endurance}; the authors emphasized that the same factors did not describe predominantly White institutions. In a longer historical analysis, \citet{ZappDahmen2025-OrganizationalMortality} connected institutional mortality to changing forms of legitimacy, including the position of HBCUs and women's colleges in an earlier period and accreditation and for-profit status in a later one. Most recently, \citet{DoughertyGao2026-SectorReligionClosure} found greater vulnerability among institutions with small enrollments and limited financial reserves, while documenting particularly steep enrollment contraction among private Catholic colleges. Read together, these findings suggest that sector, mission, special demographic focus, and religious affiliation condition exposure to market and financial pressure; they should not be read as causes of closure on their own, and their associations vary across periods and comparison groups.

Contemporary early-warning work places particular weight on trajectories rather than on a single balance-sheet value. \citet{Zemsky2020-CollegeStressTest} organize market stress around trends in new-student enrollment, net price, retention, and external financial support. \citet{Kelchen2020-EmpiricalPrediction} finds that sharp declines in enrollment, revenue, and endowment, rising tuition discounting, and federal monitoring measures are associated with subsequent closure, but also shows that most institutions classified as high risk remained open over the prediction window. This imprecision is not unique to research models: the federal financial-responsibility composite score missed a substantial share of closures and did not incorporate liquidity, historical trajectories, or forward-looking measures \citep{GAO2017-FinancialMonitoring}. Using a much richer institution-year panel, \citet{KelchenRitterWebber2024-PredictingClosures} show that revenue and expense patterns, revenue mix, operating margins, liquidity and leverage, enrollment and staffing changes, and prior signs of distress jointly improve prediction. They also find that ratios and recent changes often carry more information than levels, and that missing data are themselves unusually common among institutions that later close.

Across this literature, the recurring domains are therefore market demand and enrollment trajectory; financial resources, liquidity, leverage, cost structure, and revenue dependence; institutional scale, sector, mission, and location; regulatory and accreditation status; and the surrounding demographic and economic environment. The literature does not establish a universal checklist or a single causal pathway. Predictors differ by sector and outcome definition, governance and other unobserved events remain important, and false-positive classifications carry practical and ethical consequences. We use these domains to organize the construction of our database and to interpret the factors represented in it. Because the present study contains institutions that closed or merged but no institution-level comparison group that remained open, we do not use these variables to estimate or validate a closure-prediction model.

Here, we ask the following questions about institutions that closed or merged during this period and consider what, if anything, they suggest about the enrollment cliff, bearing in mind the distinction drawn above: 
\begin{itemize}
\item What regional trends exist in the set of closed colleges?
\item What shared characteristics do we see among closed institutions?
\item What did media messaging say about these closures? 
\end{itemize}

%------------------------------------------------

\section{Data \& Methodology}

\subsection{Dataset}

Our primary dataset integrates institutional-level characteristics with state- and regional-level socioeconomic indicators. We reviewed the empirical and applied closure literature summarized above before finalizing the feature domains, and constructed the database to represent its recurrent factors without reducing vulnerability to a single score. On the demand side, the data include total enrollment, degree level, undergraduate and graduate status, state two- and four-year enrollment shares, high-school graduation rates, the population aged 18--24, educational attainment, births, fertility, and region. To represent financial capacity and revenue context, we collected institution-level endowment values, state appropriations per full-time-equivalent student, net tuition revenue per full-time-equivalent student, personal income, and unemployment. Institutional structure and mission are represented through public or private control, nonprofit status, Carnegie classification, urban or rural location, religious affiliation, and special-demographic mission. We also distinguish outright closures from mergers and record available characteristics of the receiving institution, preserving a distinction emphasized in the closure literature. Where the sources allowed, the underlying database retains repeated annual measures for endowment, appropriations, net tuition revenue, personal income, and college-age population rather than only a single cross-section. These fields represent domains associated with vulnerability in prior work; their inclusion is not an assertion that any one of them caused the closures in our sample \citep{BatesSanterre2000-ClosuresMergers,PorterRamirez2009-WhyCollegesFail,Zemsky2020-CollegeStressTest,Kelchen2020-EmpiricalPrediction,KelchenRitterWebber2024-PredictingClosures,DoughertyGao2026-SectorReligionClosure}.

Institutional data were primarily sourced from \cite{IPEDS2025} via \cite{NCESCollegeNavigator}, providing records on enrollment, location classifications, program types, and institutional attributes. Religious affiliations and special-demographic missions were collected from primary institutional sources where applicable. To capture the pre-college student pipeline, we incorporated state-level Adjusted Cohort Graduation Rates (ACGR) from \cite{NCES-ACGR2022}.

Financial data, such as endowment information, posed a challenge. Many private institutions do not publicly disclose these figures. To address this gap, we supplemented our dataset with endowment data primarily from \cite{NTSE2025} and \cite{DataUSA2025}. Institutions were further categorized using the Carnegie Classification of Institutions of Higher Education from the American Council on Education (ACE) \cite{ace2025}.
To contextualize institutional data within broader socioeconomic environments, we integrated the following state- and regional-level indicators:
\begin{itemize}
    \item regional GDP and personal income from \cite{bea2024};
    \item state unemployment rates from \cite{bls2024};
    \item natality data, including birth rates, from \cite{cdc2024}; and
    \item general socioeconomic and demographic data from \cite{DataUSA2025}.
\end{itemize}

\paragraph{Closure identification and news corpus.} To identify institutions that closed or merged between 2020 and 2025, we began with a systematic search of news articles in 2022, supplemented by Google News alerts for ongoing updates. Once a closure was recorded, we collected institutional characteristics from \cite{NCESCollegeNavigator}. The final dataset comprises 65 institutions that closed or merged during this period, each paired with its respective institutional, state, and regional data.
To capture the qualitative dimension and public discourse surrounding closures, we also compiled and analyzed a corpus of news articles reporting on institutional shutdowns, primarily sourced from Higher Ed Dive. We used these articles to provide basic labels for the reasons for closure using the terms {\it Financial}, {\it Pandemic}, {\it Enrollment}, {\it Accreditation Issues}, and {\it Mutual Benefit} (specifically for merged institutions). These label names were taken from \cite{castillo_welding_closed_2025} but were reviewed by the researchers.

\subsection{Limitations of the Data}

A comprehensive dataset is central to this kind of analysis, and ours has several limitations that shape our methodology and the interpretation of our results. A first challenge is the nature of institutional data reporting. Although we used authoritative sources, not all data points are reported or released annually, which creates temporal gaps and limits a fully granular, year-over-year analysis for every variable. Institutions in significant distress or at imminent risk of closure are also often inconsistent in their reporting or withhold certain financial or enrollment data, so information tends to be scarcest for precisely the institutions of most interest.
 
A related challenge emerged with the collection of endowment data. Endowment figures are often complex to source, requiring careful imputation or approximation from various reports, such as those from \cite{NTSE2025} and \cite{DataUSA2025}. While these efforts were made with great care, the inherent difficulty of obtaining precise, real-time endowment values for every institution is a limitation that could be addressed in future work with more direct data streams.

The alignment with prior work is necessarily incomplete. We do not have consistent institution-year histories for enrollment, entering-cohort size, retention, tuition discounting, staffing, or expenditures, nor do we have comparable measures of operating margin, unrestricted net assets, days cash on hand, debt and leverage, federal financial-responsibility scores, Heightened Cash Monitoring, or pre-closure accreditation status. Several of these measures are among the strongest indicators in contemporary prediction studies \citep{Zemsky2020-CollegeStressTest,Kelchen2020-EmpiricalPrediction,GAO2017-FinancialMonitoring,KelchenRitterWebber2024-PredictingClosures}. Their absence means that our dataset represents the literature's major domains only partially, and is another reason to interpret the analyses as descriptions of variation among closed or merged institutions rather than as a test of which colleges are likely to fail.

Perhaps the most consequential limitation, and the one that most shaped our methodological choices, was the high dimensionality of the dataset. Our initial compilation produced a matrix of 65 institutions and 178 features. This breadth offered a wide view but posed problems for analysis, particularly for clustering, which can be affected by noise from redundant or highly correlated features. To address this, we reduced the feature set, manually removing redundant variables and working to limit dependence across features drawn from distinct sources. Silhouette scores improved after this reduction relative to the higher-dimensional version.

Our approach was designed to be as comprehensive as possible, but one strength of our data collection---its reliance on reliable governmental and scholarly sources such as IPEDS \cite{IPEDS2025}---also serves as a limitation. These sources provide a stable, consistent foundation, but they may not capture all of the qualitative or unstructured data points that influence institutional outcomes. The inclusion of a news article corpus was a deliberate attempt to address this, but it underscores the fact that no single dataset can perfectly capture the complexity of this issue.

A further limitation concerns the scope of the sample. Our dataset consists only of institutions that closed or merged during the study period; it does not include a comparison group of institutions that remained open. Constructing such a dataset was not feasible here. The full population of U.S. postsecondary institutions is large, and much of our institutional data was assembled by hand from primary sources rather than drawn from a single automated feed, which made scaling collection to all institutions impractical within the scope of this study. This has a direct bearing on how the results should be read. The Bayesian models in Section 3 do incorporate the full counts of institutions per state and region, so their estimated closure probabilities are defined relative to all institutions in a given area. The remaining analyses---the discriminant analysis of closure reasons, the principal component analysis, the clustering, and the SHAP attributions in Sections 4 and 5---are computed on the closed institutions alone. They therefore describe variation \emph{among} institutions that closed, and the patterns they surface should not be interpreted as characteristics that distinguish closed institutions from those that survived. Separating closed from surviving institutions would require a matched comparison sample, which we identify as the most important direction for future work.

Several related constraints follow from the size of the sample, and we note them collectively here in addition to the relevant points in the text. With 65 institutions, the segmentation into 21 clusters leaves only a few institutions per cluster, so individual cluster profiles are illustrative rather than stable estimates. The separation produced by the discriminant analysis (Section 4.1) is partly a consequence of having many features relative to observations, a setting in which predefined groups tend to separate regardless of their substantive meaning; we therefore read that separation descriptively. Similarly, the number of principal components needed to capture most of the variance (Section 5.1) is influenced by the closeness of the number of features and the number of observations, and the apparent tightness of the groups in the t-SNE projection partly reflects that points are colored by their assigned cluster. We flag these points so that the analyses are understood as an exploratory description of a small, hand-assembled sample rather than as precise or generalizable estimates.

\subsection{Analytical Framework}

Our aim is descriptive: to characterize the institutions that closed and the contexts in which they did so, rather than to predict future closures.

We compiled counts of the total number of higher education institutions and the number of closures per state and region to compute a baseline rate of closure. These aggregate data, independent of specific institutional characteristics, were used to construct a series of Bayesian models that estimate the probability of closure at a broad geographic level, with explicit uncertainty. Examining risk at the state and regional level also lets us revisit Grawe's argument about the role of geographic location \cite{Grawe2018-Book}.

To examine how institutional characteristics differ across the stated reasons for closure, we applied Linear Discriminant Analysis (LDA) to the closed institutions, using their officially reported closure reason as the grouping variable.  

K-means clustering was also applied to institutions to group them into segments based on shared characteristics, allowing us to describe vulnerability as something tied to an institution's specific profile rather than by the often simplistic labels assigned to them.

To add a qualitative dimension, we applied topic modeling using Latent Dirichlet Allocation to a corpus of news articles, extracting recurring themes in the public discussion of these shutdowns. We refer to the latter throughout as topic modeling to avoid confusion with the discriminant analysis.

This sequence, from broad risk estimates to a more detailed description of characteristics and narratives, is intended to be exploratory and to make the limitations of each step explicit.

%------------------------------------------------

\section{State-Level Risk Assessment}
 
\subsection{Quantifying Closure Risk}
 
The assessment of institutional closure risk at the state level was conducted through a series of increasingly complex Bayesian models, each designed to capture different facets of the underlying probability of institutional closure. The core goal of this approach is to move beyond frequentist point estimates by providing full posterior distributions for the probability of closure, thereby offering a more comprehensive understanding of uncertainty. This methodology is particularly well-suited for situations with varying data availability across states and regions, allowing for information sharing through hierarchical structures in a relatively small dataset.
 
\begin{figure}[htbp]
    \centering
    \includegraphics[width=\linewidth]{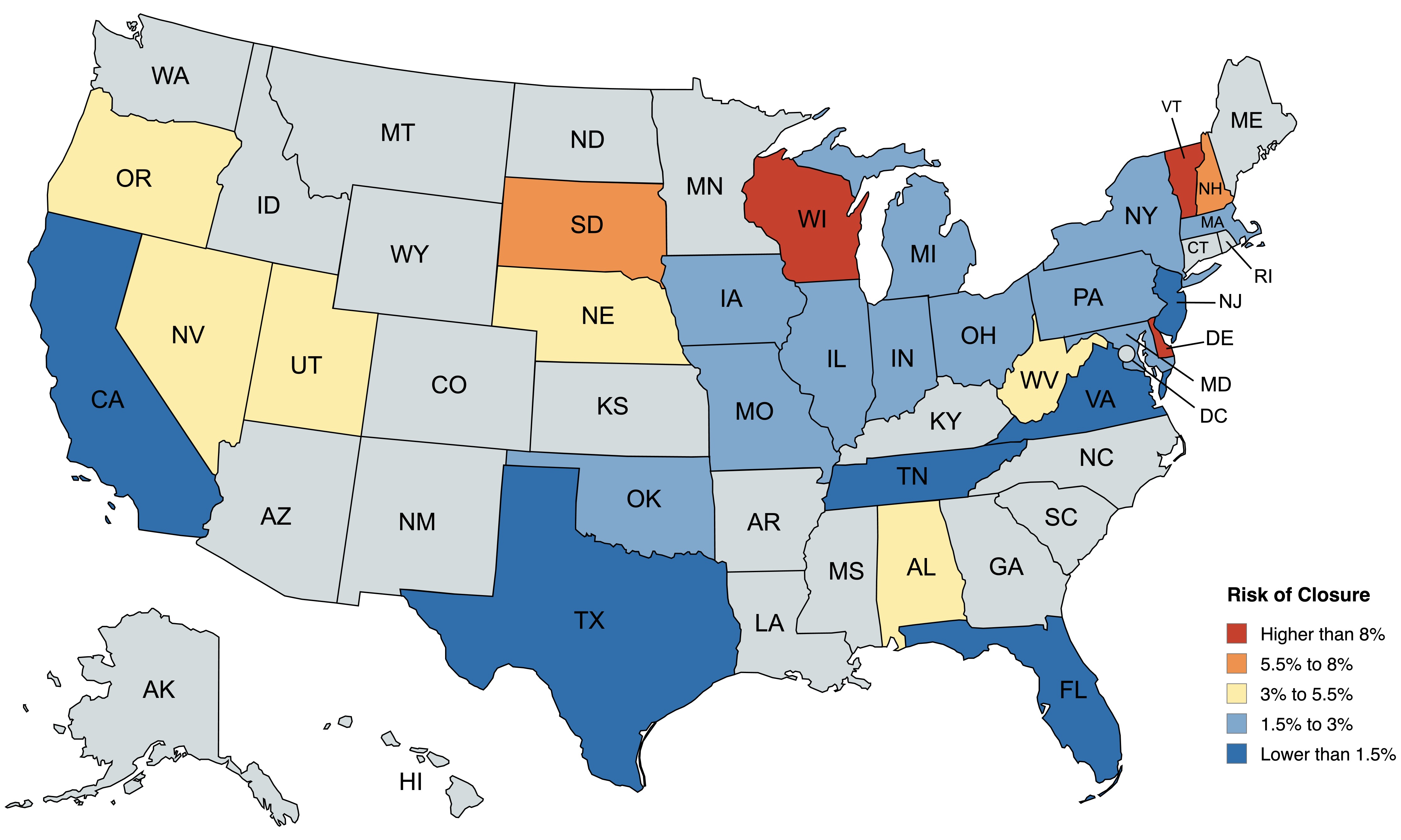}
    \caption{\footnotesize State-level estimates of institutional closure risk derived from a Bayesian model with an uninformative \(\text{Beta}(1,1)\) prior, representing uniform uncertainty over the probability of closure.}
    \label{fig:bm_beta1_state}
\end{figure}

The first model was a non-layered Bayesian model. For each state and region, the true proportion of closures, \(p\), was assumed to follow a Beta distribution, a conjugate prior for the binomial likelihood. A uniform prior, \(p\sim\text{Beta}(\alpha=1,\beta=1)\), was initially used, which assumes all values of \(p\) are equally likely before observing any data. The posterior hyperparameters were then updated as \(\alpha_{post}=\alpha + y\) and \(\beta_{post}=\beta+n-y\) where \(n\) is the total number of institutions in a given state or region and \(y\) is the number of observed closures. The state-level results of this model, which provide a baseline assessment of closure risk, are shown in Figure \ref{fig:bm_beta1_state}.

\begin{figure}[htbp]
    \centering
    \includegraphics[width=\linewidth]{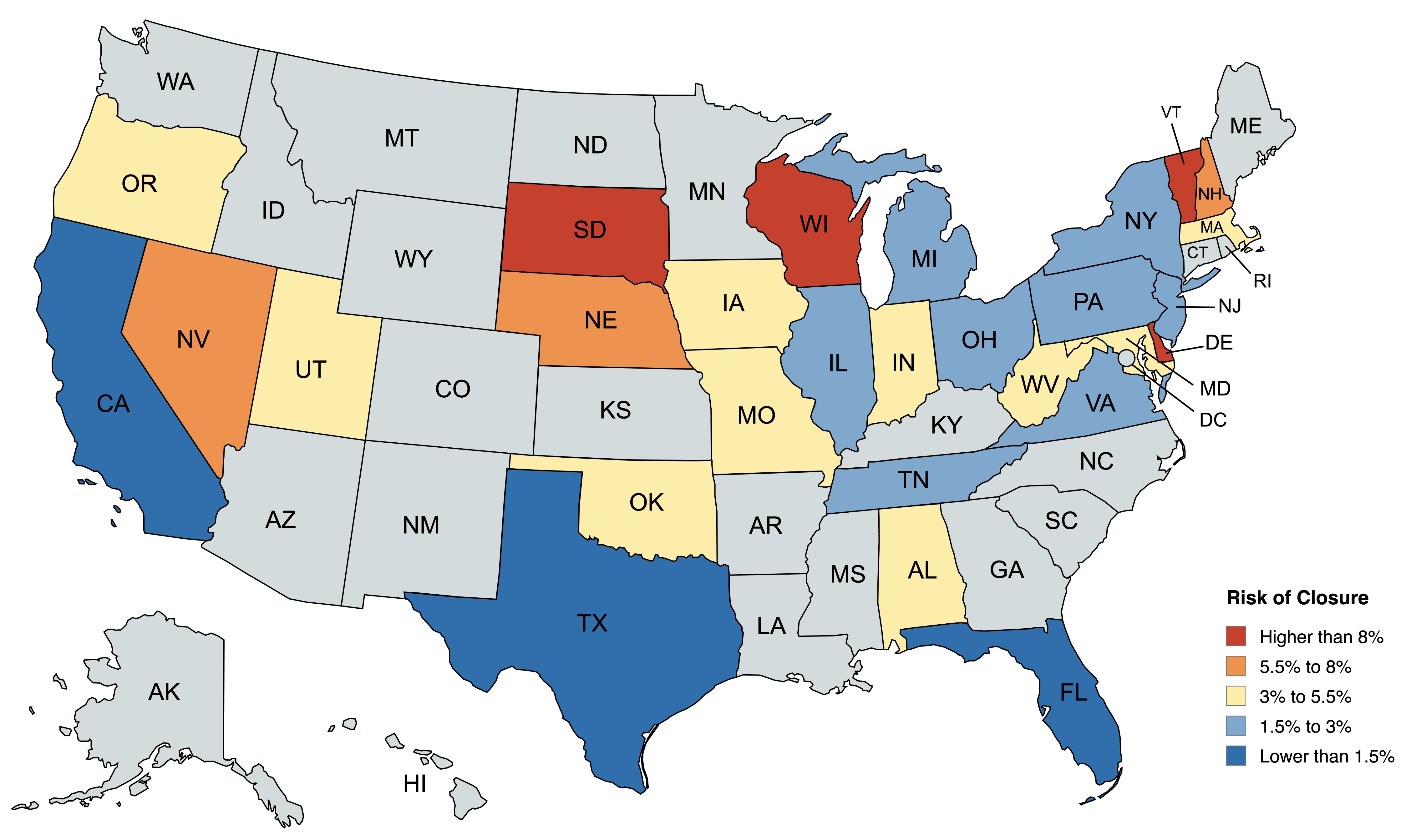}
    \caption{\footnotesize State-level estimates of institutional closure risk derived from a Bayesian model with a \(\text{Beta}(2,2)\) prior, which---unlike the uniform \(\text{Beta}(1,1)\) prior---gently centers estimates of closure around 0.5, reflecting weak regularization toward moderate closure probabilities.} 
    \label{fig:bm_beta2_state}
\end{figure}

Recognizing the influence of qualitative claims within the higher education discourse, such as the assertion that about half of institutions may be at risk of bankruptcy in the coming years from \cite{Christinen2017-HalfWillGoBankrupt}, a second non-layered model was fit using a more informative prior. This model, with \(p\sim\text{Beta}(\alpha=2,\beta=2)\), places more weight on moderate probabilities of closure, allowing the prior to reflect a stronger initial belief drawn from external claims. The results are shown in Figure \ref{fig:bm_beta2_state}. Comparing the two priors illustrates how sensitive the estimates are to prior assumptions when the data are sparse.

Subsequently, to allow for information sharing between states and regions, a simple Bayesian Hierarchical Model (BHM) was constructed. In this model, the probability of closure for each state or region was modeled as $p\sim\text{Beta}(\alpha,\beta)$, where the hyperparameters \(\alpha\) and \(\beta\) themselves were given weakly informative \(\text{Gamma}(0.01,0.01)\) priors. This hierarchical structure allows the data from all states and regions to collectively inform the estimation of the hyperparameters, producing more stable estimates for \(p\), especially in areas with few observed closures. The model was fit using Markov Chain Monte Carlo (MCMC) sampling, with 4000 samples and 2000 tuning samples per chain. Finally, a more flexible BHM was developed, also fit by MCMC, that incorporates a latent variable \(x\) and a scalar \(k\) to capture additional structure in the Beta parameters. The model includes a latent variable \(x_s\) for each state or region \(s\), drawn from a normal distribution \(x_s \sim N(\mu, \sigma)\). The hyperparameters of the Beta distribution, \(a_s\) and \(b_s\), are derived from this latent variable as \(a_s = \exp(x_s)\) and \(b_s = \exp(x_s \cdot k)\), where \(k\) is a scalar drawn from a Half-Normal distribution, \(k \sim \text{HalfNormal}(0, 0.5)\). This structure permits a more flexible representation of the processes underlying closure probabilities across states and regions. 

Moving from simple non-layered models to hierarchical structures lets us see how the estimates change as assumptions are added, and the use of full posterior distributions rather than point estimates keeps the considerable uncertainty in these small counts visible throughout.

\subsection{State and Regional Risks \& Implications}

\begin{figure}[htbp]
    \centering
    \includegraphics[width=\linewidth]{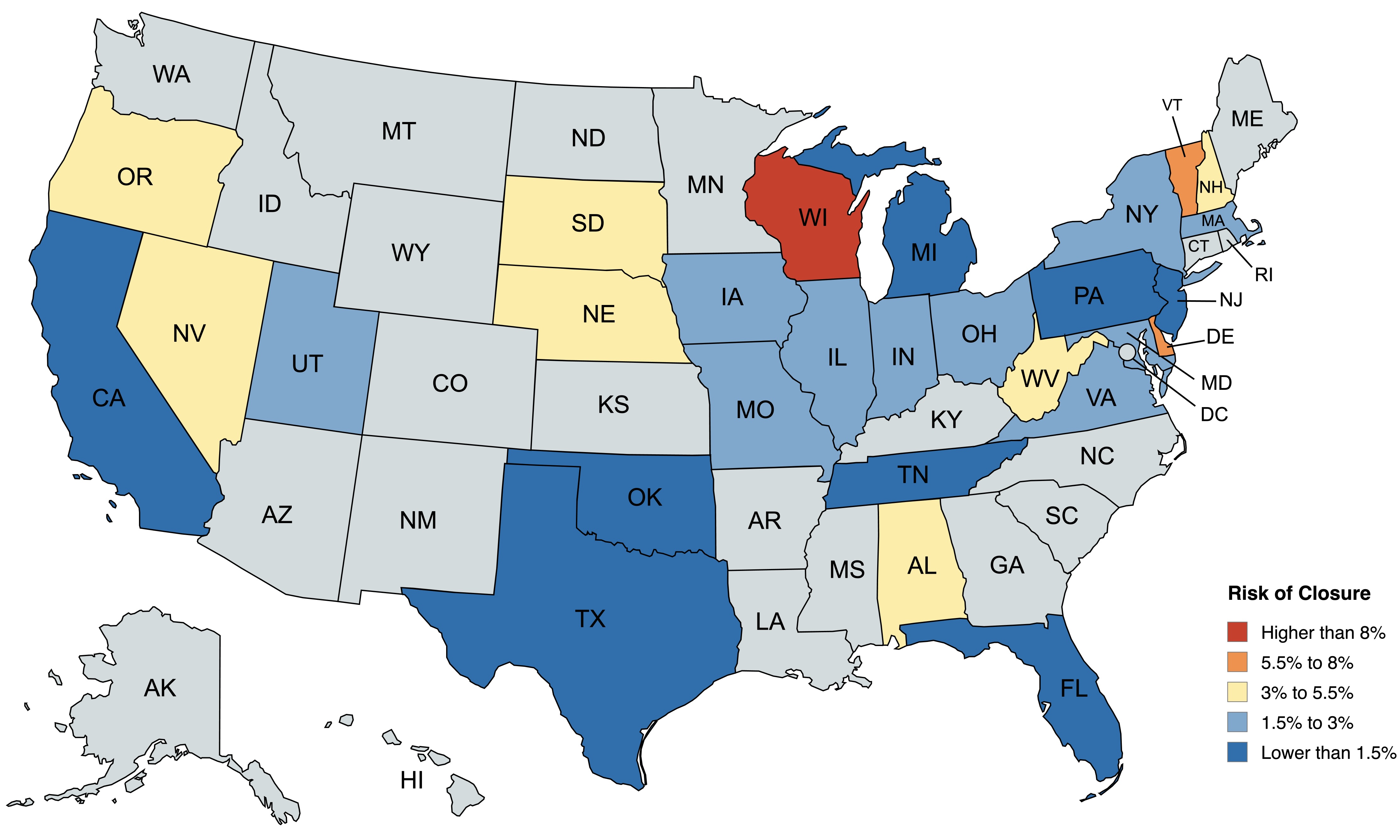}
    \caption{\footnotesize State-level estimates of institutional closure risk derived from a Bayesian hierarchical model.}
    \label{fig:bhm_state}
\end{figure}

The Bayesian hierarchical model summarizes closure risk in probabilistic terms, moving beyond a raw count of closures to estimates that carry explicit uncertainty. Figure \ref{fig:bhm_state}, generated from our four-layer BHM, shows the estimated posterior mean risk of closure for each state, grouped into risk tiers.

One feature of the hierarchical approach is the ``shrinkage'' it produces, which yields less volatile estimates for states with few institutions or few observed closures. This is visible when comparing models. In the non-layered model with a \(\text{Beta}(1,1)\) prior, the estimated risk of closure was 9\% for Wisconsin, 11\% for Delaware, and 0.5\% for Texas. With the more informative \(\text{Beta}(2,2)\) prior, these shift to 10\% for Wisconsin, 15\% for Delaware, and 0.6\% for Texas, reflecting the prior. Under the full hierarchical model, they move to 8\% for Wisconsin, 7\% for Delaware, and 0.3\% for Texas. The hierarchical model has ``borrowed strength'' from the collective data, pulling the more extreme estimates toward the overall pattern. This makes the estimates less sensitive to small sample sizes, though it does not remove the underlying uncertainty.

\begin{figure}[htbp]
    \centering
    \includegraphics[width=\linewidth]{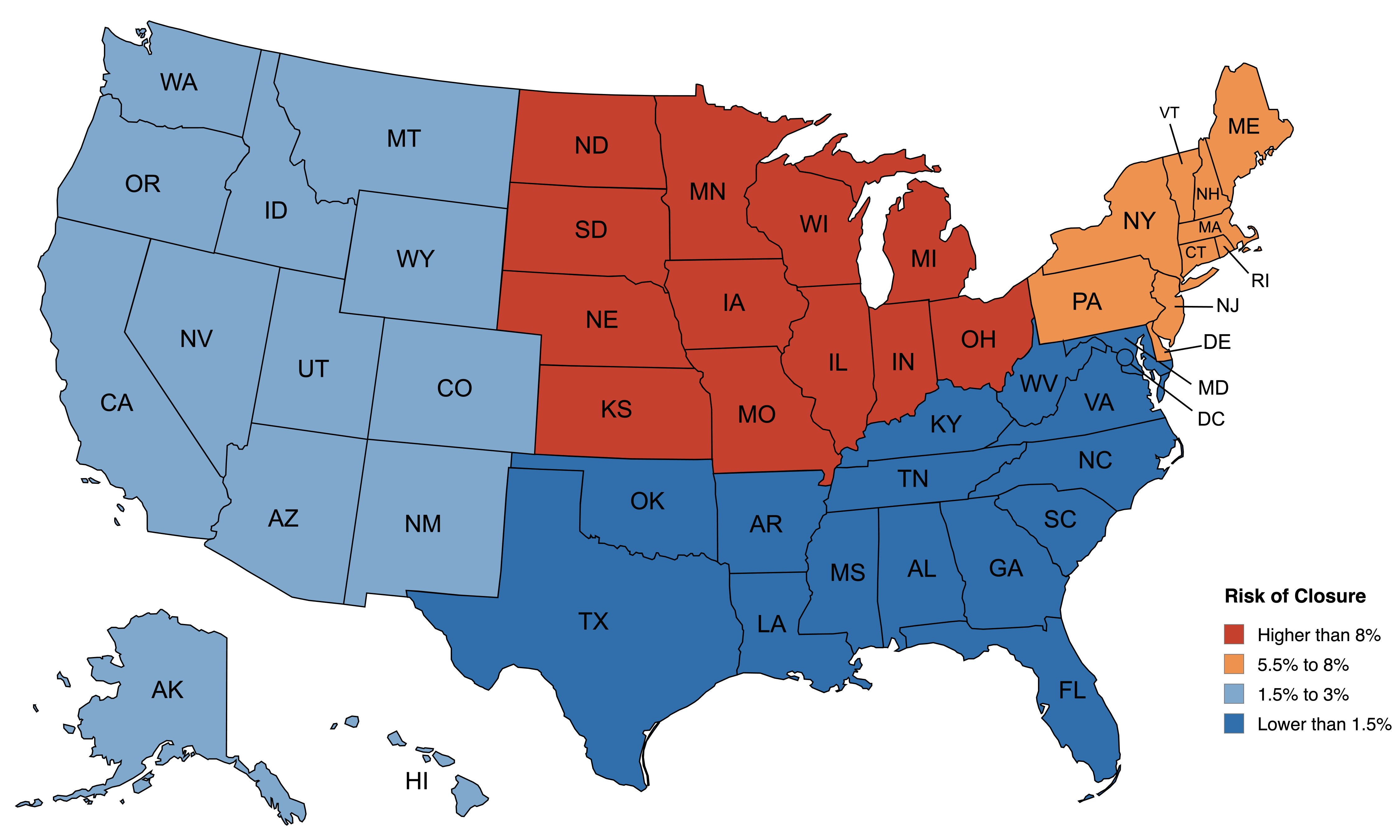}
    \caption{\footnotesize Regional-level estimates of institutional closure risk derived from a Bayesian hierarchical model.}
    \label{fig:bhm_region}
\end{figure}

The geographic pattern is broadly consistent with claims in the literature. As seen in Figure \ref{fig:bhm_state}, the higher-risk states (above 5.5\%) are concentrated in the Northeast and Midwest, including Wisconsin, Vermont, New York, and Pennsylvania. This is consistent with the projection in \cite{Grawe2018-Book} that these traditional centers of higher education would be among the more exposed to demographic decline. We read this overlap cautiously. Because the closures we observe were precipitated by the pandemic rather than by demographic decline, a shared geographic signature does not by itself confirm Grawe's mechanism: the Northeast and Midwest concentrate many small, tuition-dependent private institutions that were also among the most exposed to the pandemic shock, so the two explanations are difficult to separate in these data. Larger, more populous states in the South and West, such as California and Texas, show lower estimated risk (below 3\%). The regional-level analysis (Figure \ref{fig:bhm_region}) shows the same ordering: estimated closure risk is approximately 2\% in the Midwest and 1.5\% in the Northeast, against roughly 0.6\% in the South and 0.8\% in the West. We caution that these differences are likely tied to the number of institutions and the population base in each region, and that the absolute closure counts behind them are small.

These estimates may be of some use to policymakers and institutional leaders as a rough, probabilistic indication of where closures have been relatively more common, rather than as a forecast. They indicate that pandemic-era closures were not distributed uniformly across all regions. To the extent that the enrollment cliff pressures institutions in similar ways, its effects may also prove uneven---but our data speak directly to the pandemic period, not to the cliff itself.

We treat this as a starting point for the more granular description that follows, not as a basis for action on its own.

\subsection{The Shortcomings of Generalized Risk}

While the Bayesian models give a broad view of closure probabilities, their aggregate nature limits how useful they are for any single institution. The models assign one risk estimate to an entire state or region, implying that every institution within that boundary faces a similar level of threat. A small, rural liberal arts college in Wisconsin is assigned the same probability as a large, well-endowed public university in Milwaukee, simply because they share a state. This is reasonable for describing regional patterns, but it is of little use to a college president or board of trustees facing institution-specific circumstances.

A state-level estimate, then, does not translate into guidance for a particular institution. It does not capture the characteristics that can mitigate or worsen risk. To say anything about an individual institution, one needs a more granular description: not its risk in the abstract, but which features of the institution and its environment are associated with closure. Without that, the regional estimates remain a coarse signal rather than a guide.

This motivates the next part of our analysis. The remaining sections move from aggregate probability toward the specific characteristics and narratives associated with closure, in an attempt to describe the closed institutions in more concrete terms.

%------------------------------------------------

\section{Narratives of Closure}

\subsection{Closure Rationales}

Beyond generalized risk, a granular understanding of institutional closure requires a close examination of the officially reported reasons for shutdown. While these rationales often fall into broad categories such as financial- or enrollment-driven, our analysis reveals that the underlying characteristics of institutions that fall into these groups are fundamentally distinct. To dissect these stated rationales and reveal their unique contexts, Linear Discriminant Analysis (LDA) was applied to the subset of institutions that closed, using their officially reported primary closure reason as the grouping variable.

\begin{figure}[h]
    \centering
    \includegraphics[width=0.9\linewidth]{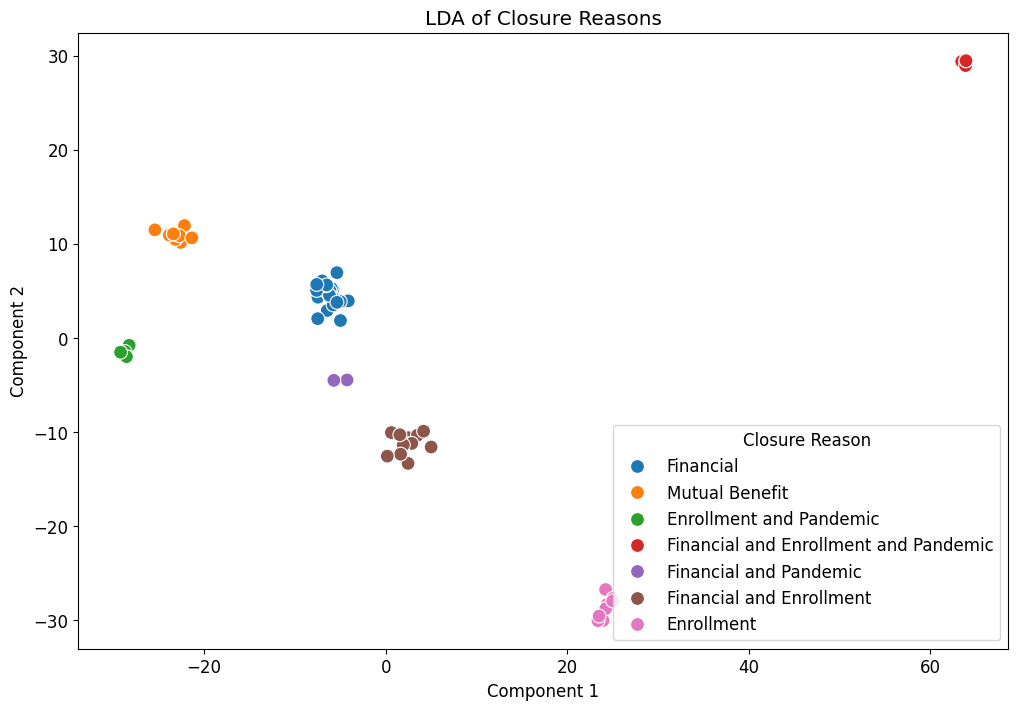}
    \caption{\footnotesize First two components of Linear Discriminant Analysis (LDA) embedding of institutional features, with points colored by closure reason. The separation among the seven closure-reason groups in the discriminant space demonstrates that institutional characteristics contain discriminative signal for predicting different closure causes.}
    \label{fig:lda_reasons}
\end{figure}

The LDA projection separates the closure-reason groups in the lower-dimensional space, as shown in Figure \ref{fig:lda_reasons}. The explained variance ratios for the discriminant components are 66.34\% for the first, 26.49\% for the second, and 5.04\% for the third, so most of the separation is carried by the first two components. We read this descriptively rather than as strong evidence: with a small sample and a comparatively large feature set, discriminant analysis will tend to separate predefined groups, so the clean separation in Figure \ref{fig:lda_reasons} should be interpreted with that caveat in mind. With that caution, the projection is consistent with the idea that institutions reporting different closure reasons differ in their broader characteristics rather than in the stated reason alone.

The analysis of the coefficients and their magnitudes for each discriminant component reveals the specific features most influential in defining these closure categories. The first component, which accounts for the majority of the discriminative power, is primarily characterized by a strong negative loading from state-level population and a strong positive loading from the proportion of the state's population that is 25 and older with some college education. This component is also positively driven by institutions offering professional and high-intensity programs, such as doctoral-only institutions, seminaries, or research-intensive universities. In contrast, the second component has a reverse relation with all of these variables, with business-school indicators also contributing. This clear opposition suggests that the components define a spectrum of institutional vulnerability, where one end is characterized by closing institutions in less-populated states with educated adult populations and professional-program specializations, while the other end represents the opposite profile. Critically, this does not highlight the significance of a low college-age population. Instead, it points to a more nuanced issue: a critical misalignment between a state's demographic profile and the demand for specific types of higher education, particularly more professional programs, which ultimately drives institutional closure.

Taken together with the separation in the LDA plot, the loadings let us describe the closure-reason groups in terms of their institutional, demographic, and socioeconomic characteristics rather than the labels alone. This is a more differentiated picture of ``college closures'' than a single category would give, though, as noted, it remains descriptive and limited by the size of the sample.

\subsection{Media Narratives}

Official statements of closure reasons are useful but often terse. To look at how closures were discussed publicly, we compiled a dataset of news articles covering each college's closure, primarily from Higher Ed Dive, and applied topic modeling (Latent Dirichlet Allocation) to identify recurring topics and themes.

\begin{figure}[h]
    \centering
    \includegraphics[width=0.85\linewidth]{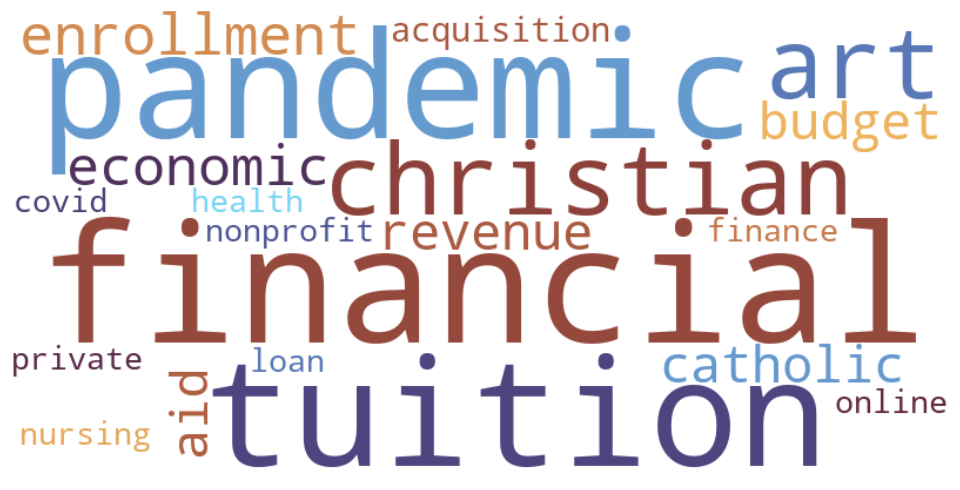}
    \caption{\footnotesize Word cloud of the top 20 terms from one of four topics identified by Latent Dirichlet Allocation on a corpus of media articles.}
    \label{fig:sample_topic}
\end{figure}

A recurring pattern across the derived topics was the co-occurrence of terms related to institutional finances and revenue with terms indicating religious affiliation. This suggests that, in the coverage we examined, financial viability and religious identity frequently appear together as themes. Figure \ref{fig:sample_topic} shows a sample topic, in which terms such as ``financial,'' ``tuition,'' ``revenue,'' ``budget,'' ``aid,'' and ``loan'' appear alongside ``Christian'' and ``Catholic.'' Other terms such as ``pandemic,'' ``enrollment,'' and ``economic'' point to the broader context in which these closures were reported.

This pattern has a few implications, offered tentatively. It suggests that institutions receiving media attention for closing often shared two characteristics: financial strain and, frequently, a religious affiliation. Where an official report might record a closure as ``financial,'' coverage tended to situate it within the institution's economic model and identity, often that of a small, private, religiously affiliated college. It also suggests that, in public discussion, the financial health of religiously affiliated institutions is treated as a notable concern.

We note that media coverage reflects what journalists and editors choose to report, not a representative measure of why institutions close, and it may over-represent topics that are newsworthy. With that caveat, the coverage incorporates a wider range of contributing factors, expert commentary, and community effects than official statements do, and the recurring pairing of financial and religious terms is consistent with the idea that closure is bound up with an institution's mission and identity as well as its finances, particularly for institutions with historical religious affiliations.

%------------------------------------------------

\section{Drivers \& Institutional Patterns}

\subsection{Uncovering Underlying Drivers}

To describe the characteristics that distinguish institutions, we applied Principal Component Analysis (PCA) to the reduced feature set, which lowers dimensionality while retaining most of the variance.

The first two components together account for 51.3\% of the total variance; the first 17 components are needed to reach 95\% of the variance. We read this as an indication that the institutional profile in our data is not well captured by a single factor, and that several roughly independent dimensions contribute to it. We note that with 65 institutions and a reduced set of 66 features, a relatively high effective dimensionality is partly expected, so we present this as descriptive support for using a multivariate approach rather than as a strong result. It does suggest that focusing on a single variable, such as enrollment or finances alone, would miss much of the variation.

Analysis of the feature loadings on these principal components reveals the most influential factors. The first principal component (PC1) is strongly influenced by multiple state-specific features related to population size in certain age brackets (e.g., state's total 18--24-year-old population) alongside a few financial variables. Conversely, the second principal component (PC2) is largely shaped by features such as high school graduation rates (ACGR) and regional fertility rates contributing positively, while net tuition costs and personal income figures show substantial negative loadings. The distinct contributions of these variables underscore the separation between regional and state-level demographic and educational attainment factors, on the one hand, and financial indicators, on the other, as key axes of variation among institutions.

\subsection{Segments \& Institutional Archetypes}

Building on the dimensionality reduction, K-means clustering was applied to group institutions by their characteristics. The number of clusters was set to 21, guided by silhouette score analysis. The resulting distribution across clusters shows segments of varying size. In the reduced-dimensional view, the groups appear reasonably well separated and tight, suggesting that institutions within a cluster are similar to one another. We emphasize that 21 clusters over 65 institutions leaves only about three institutions per cluster on average, and several clusters are smaller still. The clusters should therefore be read as a fine-grained, exploratory grouping rather than as stable, generalizable types, and individual cluster profiles should be interpreted cautiously.

\begin{figure}[h]
    \centering
    \includegraphics[width=0.85\linewidth]{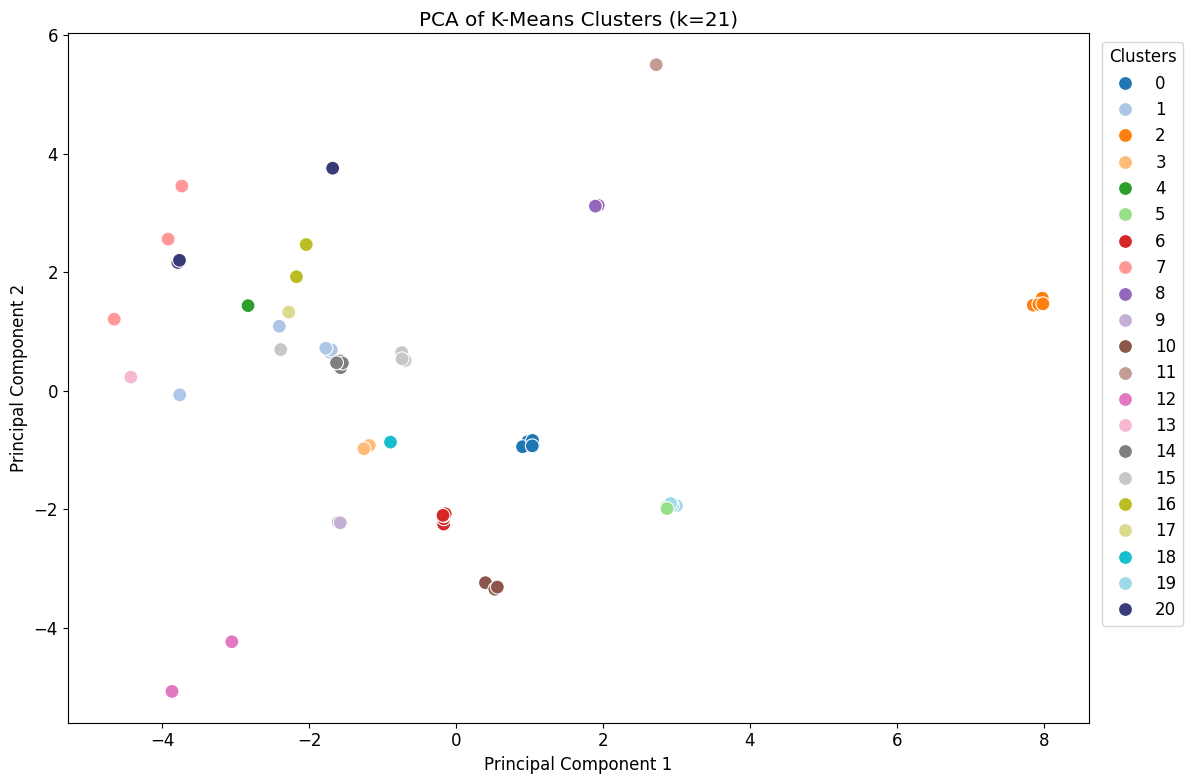}
    \caption{\footnotesize PCA with K-means clusters (k=21)}
    \label{fig:pca_clusters}
\end{figure}

To look at the structure of the data, we used t-distributed stochastic neighbor embedding (t-SNE) with labels coming from k-means clustering to visualize the groupings. The plot showed points falling into reasonably distinct groups.

\begin{figure*}[h]
    \centering
    \includegraphics[width=0.85\linewidth]{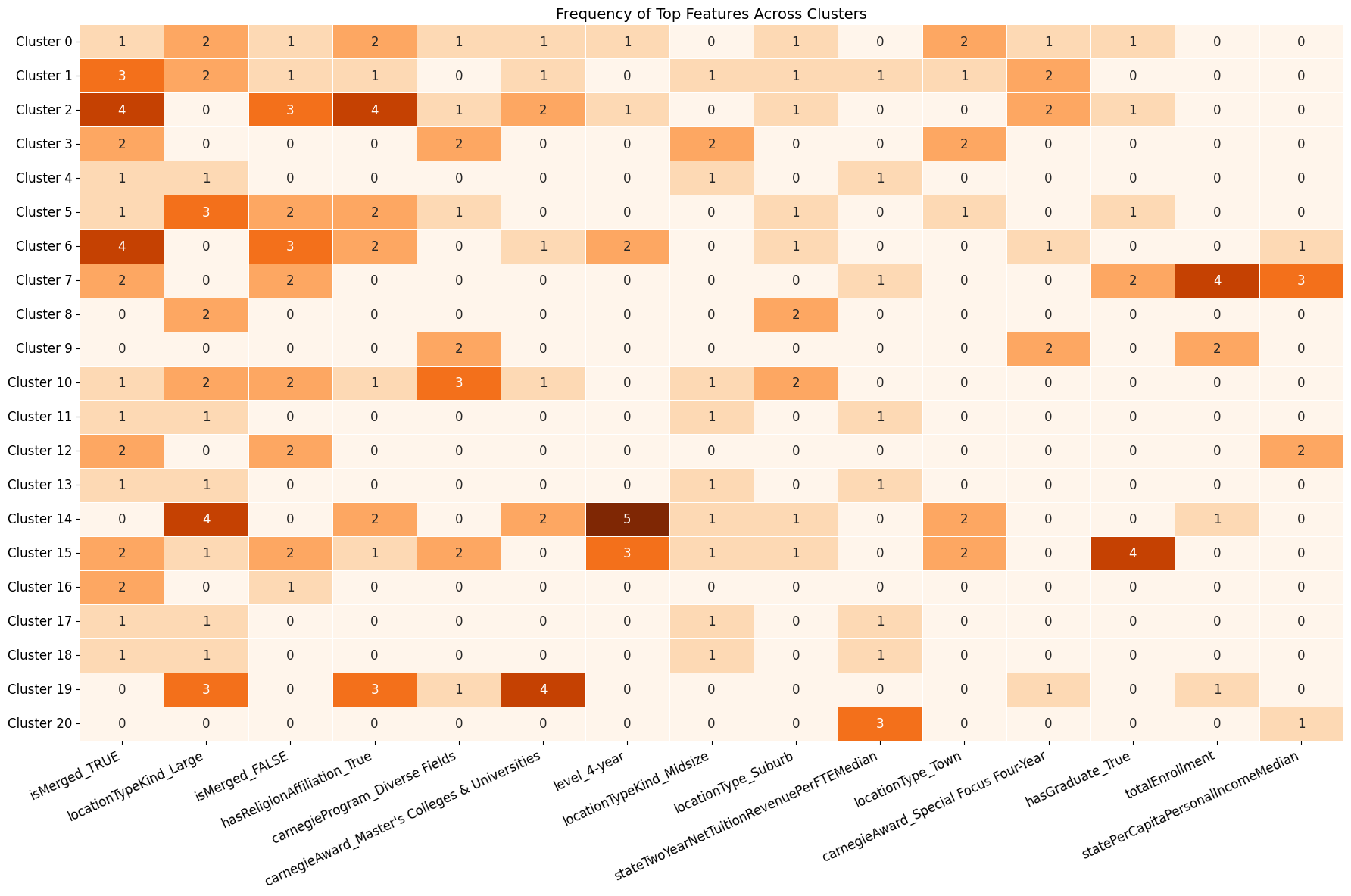}
    \caption{\footnotesize Frequency of top features across clusters}
    \label{fig:feature_frequency}
\end{figure*}

We examined the frequency of top features within each cluster (Figure \ref{fig:feature_frequency}), which shows which attributes are most common in each of the 21 groups. One cluster, for example, may be characterized by variables associated with large institutions that have recently merged, while another is marked by variables associated with four-year religiously affiliated institutions. This supports a qualitative reading of each cluster in terms of recognizable profiles, such as ``financially stable research universities,'' ``smaller liberal arts colleges under financial pressure,'' or ``community colleges serving particular regional needs,'' while keeping in mind the small number of institutions behind each label.

To describe which features are most associated with cluster membership, we computed Shapley additive explanations (SHAP) values, a measure of feature importance from cooperative game theory that quantifies how much each variable contributes to an institution's placement in a given cluster. The features with the highest mean SHAP values span financial, demographic, and socioeconomic categories, again pointing to several distinct contributing factors rather than one.
 
These features fall into three groups. The first is financial: appropriation amounts for four-year and two-year (largely community) colleges in a state, state tuition revenue, and the median endowment of colleges. This points to an institution's financial structure, and its reliance on public funding versus private wealth, as a factor in how it is grouped. The second is regional demographics: the population of the region, regional fertility rate, high school graduation rate (ACGR), and the share of the state population with some college education. These factors connect an institution's local student pipeline to its profile, linking back to the demographic shifts described in \cite{Grawe2018-Book}. The third is the broader economic environment, including state unemployment rate and per capita income. Together, these features suggest that an institution's grouping reflects a combination of financial, demographic, and economic characteristics rather than any single attribute.
 
This description points to some concrete considerations for institutions. A college whose profile is defined largely by reliance on state appropriations might consider diversifying its funding, and one defined by its regional birth rate might consider broadening recruitment to non-traditional students and students from other regions and countries. These are illustrative rather than prescriptive. More broadly, the clustering and SHAP show that the institutions that closed during this period were heterogeneous rather than uniform. If the enrollment cliff acts on institutions through similar characteristics, it too may produce an uneven, segmented pattern rather than a uniform one---though, again, this evidence is drawn from pandemic-era closures rather than from the cliff.

%------------------------------------------------

\section{Discussion \& Conclusion}

\subsection{Summary of Findings}

This study examined the institutions that closed or merged between 2020 and 2025 and the contexts in which they did so. Across the analyses, the closed institutions appear heterogeneous rather than uniform, and their circumstances combine demographic, financial, and regional factors. Framed around the three questions posed in the introduction, the results are straightforward: closures were regionally uneven, the institutions shared no single profile, and media coverage emphasized finances, enrollment pressure, and institutional identity, especially religious affiliation. We summarize how these results relate to the existing discussion, while keeping in mind the descriptive and small-sample nature of the work.
 
\cite{Grawe2018-Book} identified the demographic headwinds facing higher education, arguing that a decline in college-aged populations would create challenges for institutions. Our hierarchical models are consistent with the geographic part of this claim, estimating relatively higher closure rates in the Northeast and Midwest. We note, however, that these models describe pandemic-era closures, not closures caused by demographic decline; the agreement with Grawe's geographic projection is therefore suggestive rather than confirmatory, and may partly reflect the concentration of pandemic-vulnerable institutions in the same regions. 

At the same time, our results are consistent with the view that COVID (and perhaps the enrollment cliff as well) is not a single phenomenon driven by birth rates alone: in our data, the institutions that closed differ along financial, operational, and regional-socioeconomic dimensions as well as demographic ones. We present this as a description of the closed institutions rather than as a causal account.
 
The Principal Component Analysis is consistent with this reading: the first two components account for just over 51\% of the variance, and 17 components are needed to reach 95\%, suggesting that a single-factor explanation would capture only part of the variation (with the small-sample caveat noted in Section 5.1). The clustering produced 21 segments, each with a different mix of high-SHAP features, which is consistent with the closures in our sample reflecting several distinct institutional profiles. If the enrollment cliff operates through the same characteristics, it may likewise affect institutions unevenly rather than uniformly. This more granular reading is in line with discussions found in \cite{Seybold2024-ControvertialTakeTweet, reddit2024-ControvertialTakeTweet}, which recommend looking beyond a single ``cliff.''
 
The closure narratives point in a similar direction. The discriminant analysis of official closure reasons is consistent with institutions that received a ``financial'' label differing in their broader characteristics from those that received an ``enrollment'' label, suggesting that a single label for institutional failure obscures real differences. The topic modeling of media coverage surfaced a recurring pairing of financial viability and religious affiliation, suggesting that the press often reads closure through an institution's identity as well as its balance sheet, particularly for smaller, private, religiously affiliated colleges. Taken together, these results offer an empirical, if preliminary, description of which institutions closed during this period and how those closures were discussed. They suggest that the picture is segmented and multi-factorial rather than uniform.

\subsection{Implications for Institutions and Policymakers}

The preceding analysis describes some of the characteristics associated with institutional closures in this period. Here we discuss what these descriptions might suggest for institutions and policymakers, offered cautiously and as directions to consider rather than recommendations. We do not claim that any institution can be made immune to demographic and financial pressure; the more modest point is that the factors we observe vary across institutions, and responses can reasonably vary with them.
 
One recurring pattern in the media coverage was the pairing of financial strain and religious affiliation. For some smaller, private, religiously affiliated institutions, a distinctive mission may coincide with financial fragility, particularly alongside declining enrollment. Consolidation or merger is one option such institutions sometimes pursue. Combining several smaller religious colleges into a larger institution could, in principle, yield economies of scale, broaden academic offerings, and pool endowment resources, while retaining a shared mission. This resembles the ``Mutual Benefit'' category that appeared among the closure reasons in our data, in which a closure is structured as a combination rather than an outright shutdown. Whether such mergers succeed depends on circumstances our data do not capture.
 
Our results also point to the role of state appropriations and endowments in the financial profiles of institutions, at a time when public funding has in many cases been under pressure. This raises questions about public and private investment in higher education that are largely outside the scope of our data, but that the closures make salient. To the extent that policymakers treat higher education as contributing to economic activity and mobility, targeted public funding for institutions serving at-risk populations or located in higher-risk regions is one lever available to them; private investment in endowments, scholarships, and program development is another. We note these as policy levers that bear on the factors we observe, without taking a position on how they should be prioritized.
 
Labor-market conditions and student expectations are also part of the context in which institutions operate. The value of a broad liberal arts education is well established, including its role in preparing students for a range of contributions to society as discussed in \cite{Lemke2009-PhdAtLibArt}, and our results do not bear on that question directly. They do sit within a broader discussion about the place of career-oriented programs. One framing in that discussion is not to replace the liberal arts with vocational training but to integrate career preparation within it; institutions that combine academic programs with career services and industry connections may be better positioned for shifts in student demand. We raise this as part of the surrounding discourse rather than as a finding of our analysis.
 
Finally, the regional estimates and cluster descriptions may be of some use in thinking about where support could be directed. Regions with relatively higher closure rates in our estimates might warrant closer attention, through mechanisms such as regional consortia for resource sharing, shared recruitment efforts, or programs aimed at non-traditional and adult learners. Shared data platforms among state institutions could support earlier identification of financial or enrollment difficulties. For an individual institution, knowing which characteristics most define its cluster (for example, heavy reliance on state appropriations versus a declining regional birth rate) may help focus planning. We offer these as considerations consistent with our descriptive results, not as a strategy that guarantees any particular outcome.

\subsection{The Ethics of Predictive Modeling}
 
Although the methods used here could be extended toward a predictive model that labels institutions as likely to close, we deliberately do not pursue that, and we are wary of such models. Beyond the small sample, building a closure-prediction label raises practical and ethical problems that, in our view, outweigh its benefits in this setting.
 
A primary ethical issue stems from the inherent imbalances and biases in the data. Institutions that are struggling are often the least capable of consistently reporting comprehensive and accurate data, creating a feedback loop where an institution's distress is correlated with a lack of data, which a predictive model might misinterpret as a definitive sign of risk. Furthermore, a predictive label of ``at risk of closure'' could itself become a self-fulfilling prophecy. Such a designation, if made public, could trigger a rapid decline in enrollment, loss of faculty and staff, and a withdrawal of donor support, accelerating the very outcome the model was designed to predict.
 
Rather than a predictive label, we have aimed to provide descriptive context that can inform decisions. We see the more useful role of these methods here as identifying the characteristics associated with an institution's profile, not as assigning a binary outcome.

A visualization of this kind, read alongside the rest of the analysis, can give an institution a sense of how it is positioned relative to its peers and which characteristics most define its profile. An institution might find that it is defined largely by reliance on state appropriations or by location in a region with a declining birth rate. That knowledge can inform locally appropriate responses, such as diversifying funding or broadening recruitment, in a way that a single generic risk score cannot. Our emphasis is on context and interpretability rather than on a predictive output, with the aim of describing rather than labeling institutions.

\subsection{Future Directions}

The most important extension is the construction of a comparison group of institutions that remained open. Because our analyses of closure reasons and institutional archetypes were computed on closed institutions alone, they describe differences within that set rather than the characteristics that separate closed from surviving institutions. A tractable path, given that much of our dataset was assembled by hand, would be to compile a matched or stratified sample of surviving institutions---comparable in size, sector, control, and region---rather than the full national population. With such a comparison group in place, the methods used here could be redirected toward identifying the features most associated with closure itself, allowing the descriptive patterns we report to be tested rather than only observed.

A useful direction for future research is a closer look at shifting enrollment demographics, particularly growing enrollment among groups that have historically been underrepresented. The enrollment cliff is often framed around declining birth rates in traditional college-going populations, but enrollment growth among other demographic segments may offset part of that decline. Examining the characteristics, needs, and enrollment patterns of these student populations, including academic preparation, financial aid needs, program preferences, and retention, would give a fuller picture of where institutions might find opportunities for growth, and which institutions are serving these students most effectively.
 
Beyond demographic shifts, future work could enhance the predictive power of our models by incorporating more dynamic, time-series data. While our current Bayesian models provide a snapshot of risk, integrating longitudinal financial, enrollment, and demographic data would enable the development of predictive models capable of forecasting closures with greater precision over various time horizons. This would involve techniques like dynamic Bayesian networks or state-space models, allowing for the real-time monitoring of institutional health indicators and the identification of early warning signals.
 
Furthermore, a deeper dive into program-level analysis is warranted. While this study focused on institutional-level characteristics, the viability of a college is often tied to the strength and relevance of its individual academic programs. Future research could explore which specific disciplines or vocational pathways are experiencing growth or decline, and how an institution's portfolio of offerings affects its overall financial stability and attractiveness to students. Such work could inform strategic program development and resource allocation decisions.
 
Finally, exploring the broader societal impacts of closures and the effectiveness of various mitigation strategies would be invaluable. This could involve qualitative research, such as case studies of institutions that successfully adapted to pressures, or those that closed and the subsequent effects on their local communities. Understanding the long-term economic and social consequences of institutional failure, and conversely, the benefits of successful adaptation, would provide a more complete picture of the challenges and opportunities facing higher education in the coming decades. By continually refining our data, methodologies, and areas of inquiry, we can better equip stakeholders to navigate the complexities of the post-2020 higher education environment and foster a more resilient and equitable system.

\section*{Data availability}

The data and code supporting the findings of this study are available from the authors upon request.

%----------------------------------------------------------------------------------------
%	REFERENCE LIST
%----------------------------------------------------------------------------------------

\begingroup
\small
\sloppy
\setlength{\bibsep}{2pt plus 0.3ex}
\setlength{\emergencystretch}{3em}
\phantomsection
\bibliographystyle{plainnat}
\bibliography{references}
\endgroup

%----------------------------------------------------------------------------------------

\end{document}